\documentclass[conference]{IEEEtran}
\IEEEoverridecommandlockouts
\usepackage{cite}
\usepackage{amsmath,amssymb,amsfonts}
\usepackage{algorithmic}
\usepackage{graphicx}
\usepackage{textcomp}
\usepackage{xcolor}

\usepackage{url}
\usepackage{subcaption}
\usepackage{soul}
\usepackage{booktabs}
\usepackage{siunitx}
\usepackage{hyperref}

\newcommand\rurl[1]{%
  \href{https://#1}{\nolinkurl{#1}}%
}

\usepackage[top=0.75in, left=0.63in, right=0.63in, bottom=1.04in]{geometry}
\def\BibTeX{{\rm B\kern-.05em{\sc i\kern-.025em b}\kern-.08em
    T\kern-.1667em\lower.7ex\hbox{E}\kern-.125emX}}
\begin{document}

\title{Accented Text-To-Speech Synthesis With a Conditional Variational Autoencoder
\thanks{This project has received funding from SUTD Kickstarter Initiative no. SKI 2021\_04\_06.}
}

\author{
    \IEEEauthorblockN{Jan Melechovsky\IEEEauthorrefmark{1}, Ambuj Mehrish\IEEEauthorrefmark{1},
    Berrak Sisman\IEEEauthorrefmark{2}, and Dorien Herremans\IEEEauthorrefmark{1}
    }
    \IEEEauthorblockA{
        \IEEEauthorrefmark{1}Audio, Music, and AI Lab, Singapore University of Technology and Design, Singapore\\
        jan\_melechovsky@mymail.sutd.edu.sg, ambuj\_mehrish@sutd.edu.sg, dorien\_herremans@sutd.edu.sg
    }
    \IEEEauthorblockA{
        \IEEEauthorrefmark{2}Speech \& Machine Learning Lab, The University of Texas at Dallas, USA\\
        Berrak.Sisman@UTDallas.edu
    }
}

\maketitle

\begin{abstract}
Accent plays a significant role in speech communication, influencing one's capability to understand as well as conveying a person's identity. This paper introduces a novel and efficient framework for accented Text-to-Speech (TTS) synthesis based on a Conditional Variational Autoencoder. It has the ability to synthesize a selected speaker's voice, and convert this to any desired target accent. Our thorough experiments validate the effectiveness of the proposed framework using both objective and subjective evaluations. The results also show remarkable performance in terms of the model's ability to manipulate accents in the synthesized speech. Overall, our proposed framework presents a promising avenue for future accented TTS research.
\end{abstract}


\section{Introduction}
\label{sec:intro}
Accent in speech involves variations in phoneme, rhythm, intonation, and structure. It carries information about a person's background such as education, region, and mother tongue \cite{wells1982accents}. Disentangling accent from other speaker characteristics like pitch and vocal tract shape has proven to be challenging, as accent is a part of one's idiolect.

Accented Text-to-Speech (TTS) has many real-world applications, but it has not been a primary focus in the TTS field. Incorporating accent into TTS models enhances customizability and improves identity representation for people with speaking disabilities. Changing the accent in conversational AI could also aid user comprehension.
We hypothesize that learning accent representations as controllable attributes with a multi-accent dataset is a more data-efficient way than training separate accent-specific systems, as the general speaking ability of the model is learnt across all the data.

Recent advancements in TTS, including attention-based deep learning models like Tacotron \cite{wang2017tacotron}, Tacotron2 \cite{shen2018natural}, FastSpeech \cite{ren2019fastspeech}, and FastSpeech2 \cite{ren2020fastspeech}, have significantly improved performance. For controllable speech synthesis, contributions include Global Style Tokens (GST) \cite{wang2018style}, GMVAE-Tacotron \cite{hsu2018hierarchical}, and Variational Autoencoders (VAE) for speaking style modeling. GST captures speech attributes from the reference audio, while GMVAE and VAE focus on latent prosody attributes like affect and intent. Recent work has also focused on disentangling speakers and accents using Multi-Level VAE~\cite{melechovsky2023learning}.

Most research on accented speech focuses on foreign accent conversion (FAC), accent identification, and accented automatic speech recognition. FAC is a type of voice conversion aiming to make L2 speech sound like L1 speech, achieved through disentanglement using pre-trained classifiers or adversarial learning. In this paper, we go beyond style transfer, and develop a TTS system capable of generating speech in any accent for any speaker without source audio.

We propose an efficient TTS system based on Tacotron2 that uses a Conditional Variational Autoencoder (CVAE) for accent conditioning while retaining speaker identity. The CVAE architecture enables controllable features such as speaker identity, emotion, or style. Our contributions include: 1) A novel framework for controllable speech synthesis with a focus on accent conversion, 2) Disentanglement of accent and speaker characteristics, 3) Accent conversion without reference audio, and 4) Elaborate evaluation with insightful discussion.

The rest of the paper is organized as follows: Section~\ref{sec:relwork} covers related work, next, in Section~\ref{sec:method} we describe our proposed method. Then, Section~\ref{sec:exp} details the training procedure and experiments, and finally, Section~\ref{sec:conc} concludes the study.

\begin{figure*}[h]
    \centering
    \includegraphics[width=1.5\columnwidth]{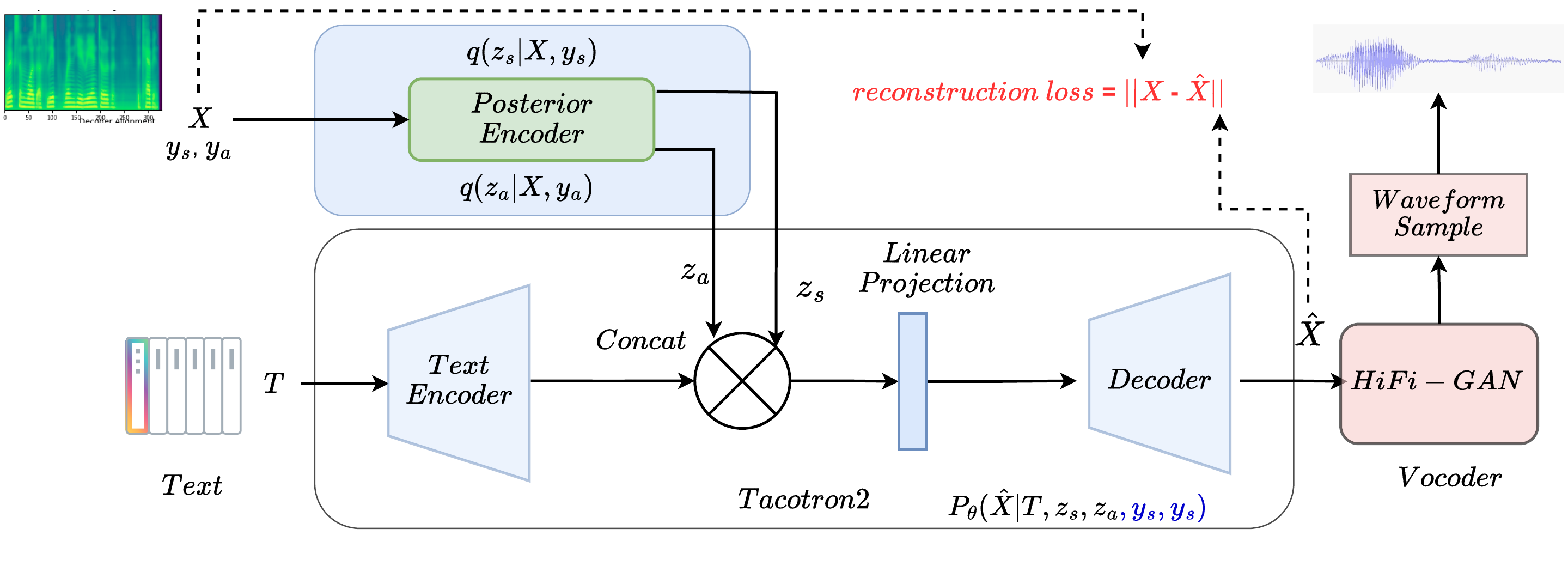}
    \vspace{-3mm}
    \caption{An illustration of the training phase and overall architecture, includes Tacotron2 with a CVAE encoder.}
    \label{fig:CVAE}
    \vspace{-5mm}
\end{figure*}

\section{Related Work}
\label{sec:relwork}

The evolution of TTS systems has advanced significantly with the introduction of Tacotron by Wang et al. \cite{wang2017tacotron} and Tacotron 2 by Shen et al. \cite{shen2018natural}, leveraging end-to-end neural networks and WaveNet \cite{oord2016wavenet} for more natural speech. FastSpeech and FastSpeech 2 by Ren et al. \cite{ren2020fastspeech} improved efficiency and reduced latency with a non-autoregressive architecture. Recent research has focused on enhancing control and expressiveness in generated speech. Architectures like GST \cite{wang2018style} and GMVAE \cite{hsu2018hierarchical}, used with Tacotron 2, offer greater controllability, enabling emotional speech generation, accent conversion, and style conversion. In our work, we use GST and GMVAE as baselines, following the architecture from \cite{hsu2018hierarchical} with two observed encoders (one for accent and one for speakers) and no latent encoder, each with an embedding size of 16. Our proposed method, uses a Conditional VAE (CVAE) to capture accent and speaker characteristics.

\section{Proposed Method}
\label{sec:method}


The architecture of the proposed method is shown in Fig.~\ref{fig:CVAE}. It consists of Tacotron2~\cite{shen2018natural} and a Posterior Encoder (Fig.~\ref{fig:PE}). As the Posterior Encoder, we have opted to use a CVAE architecture~\cite{sohn2015learning} with the objective of maximizing the evidence lower bound (ELBO) of the intractable marginal log-likelihood of data $\log p_{\theta}(X|\textbf{y})$:
\begin{figure}[h]
    \centering
    \includegraphics[scale=0.41]{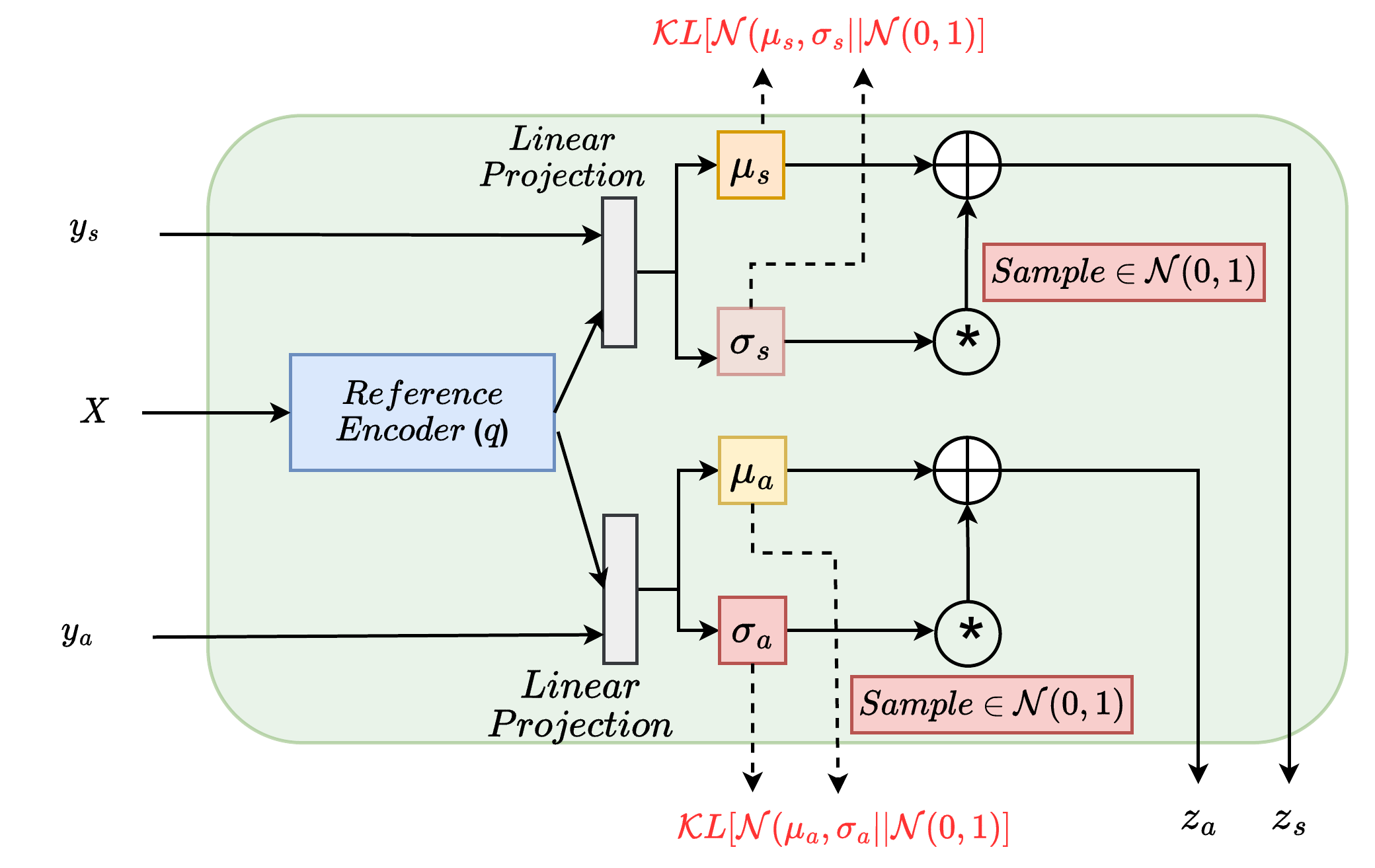}
    \caption{Posterior Encoder architecture based on CVAE.}
    \label{fig:PE}
    \vspace{-3mm}
\end{figure}
\begin{equation}
    \log p_{\theta}(X|\textbf{y}) \geq \mathbb{E}_{q_{\phi}(z|X)}[\log p_{\theta}(X|\textbf{z})-\log \frac{q_{\phi}(z|X)}{p_{\theta}(X|\textbf{y})}]
\end{equation}
 where $\theta$ and $\phi$ represent the parameters used for the decoder and posterior encoder respectively and $p_{\theta}(X|\textbf{y})$ denotes a posterior distribution of latent variable $z = [z_{s},z_{a}]$ with given label condition $y=[y_{s},y_{a}]$
 for a speaker $s$ and accent $a$.
 The negative ELBO is then used as training loss, which can be viewed as the sum of the reconstruction loss $-\log p_{\theta}(X|\textbf{z})$ and the KL divergence loss $\log q_{\phi}(z|X) - p_{\theta}(X|\textbf{y})$.

The $L_{2}$ loss between the predicted mel~spectrogram $\hat{X}$ and ground truth mel~spectrogram $X$ is used as the reconstruction loss:

\begin{equation}
    L_{recon} = ||\hat{X}-X||_{2}
\end{equation}

where $||.||_{2}$  denotes $L_{2}$ norm. For the CVAE encoder, we propose two variants. The first one follows the traditional CVAE concept of having a label passed as a condition to both the encoder and decoder. The intuition is that the speaker and accent are mainly determined by the provided labels and the latent distribution captures minor differences inside these categories, like prosody. The second variant uses labels only in the encoder. Thus, the whole accent and speaker representation is captured by the latent variables $z_a$ for accent and $z_s$ for speaker. We name these two variants CVAE-L, and CVAE-NL, indicating `label', and `no-label', respectively.




    

The generated $z_a$ and $z_s$, each of size $128$, are concatenated with the text embeddings and passed through a single linear layer. The output is then passed to the decoder to generate a mel spectrogram, which is converted into audio with a pre-trained HiFi-GAN.
\section{Experiments}

\label{sec:exp}

\subsection{Dataset}
In our experiments, we used the L2Arctic dataset \cite{zhao2018l2}, which contains 27 hours of speech from $24$ speakers with $6$ distinct accents, each represented by two female and two male speakers. We selected 10 unseen utterances per speaker for testing, 15 seen and 5 unseen utterances per speaker for validation, and the remaining data for training.

\begin{figure}[ht]
    \centering
    \begin{subfigure}{\columnwidth}
        \includegraphics[width=\columnwidth,height=0.45\columnwidth]{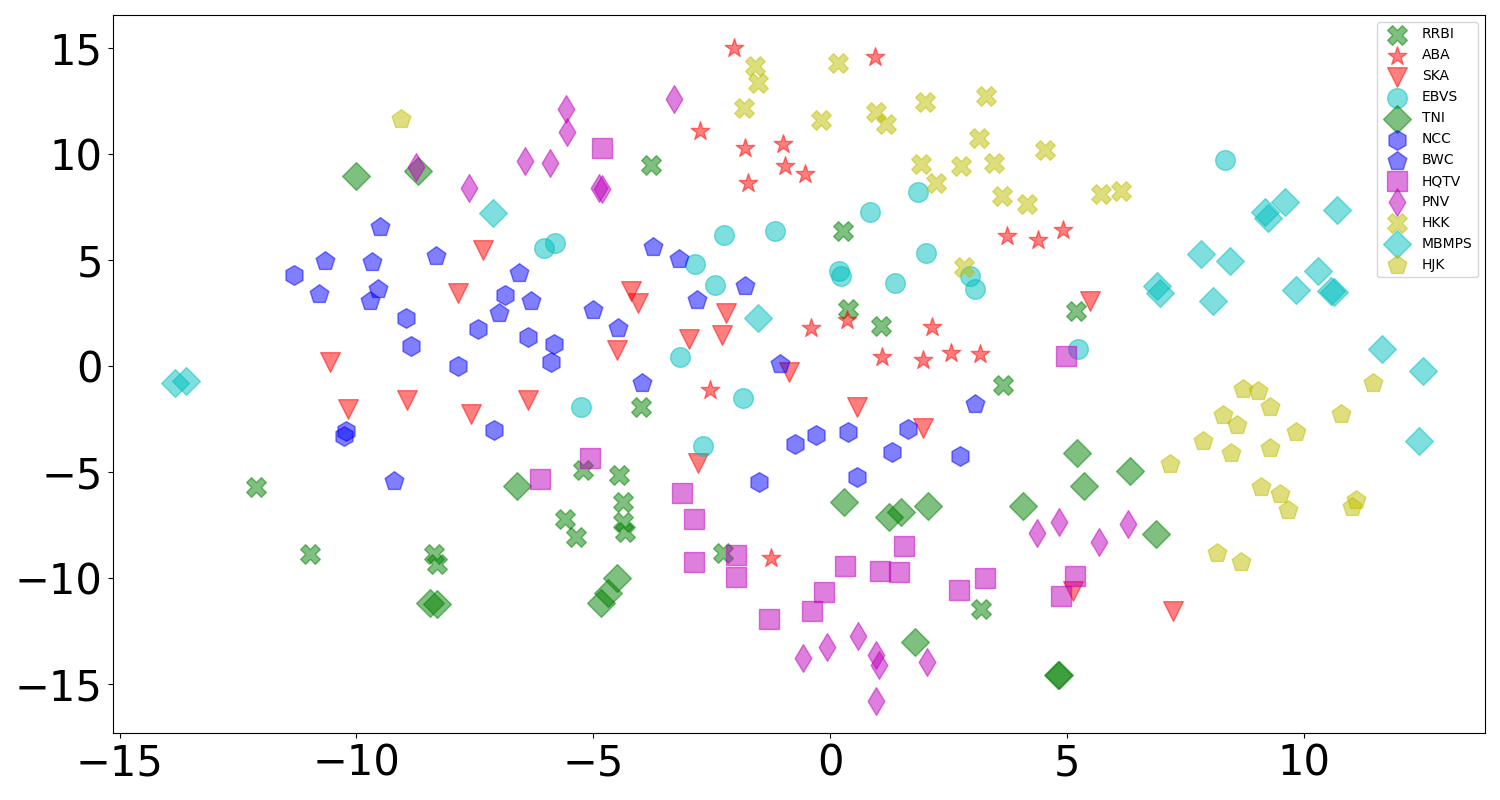} 
        \vspace{-5mm}
        \caption{Speaker embeddings $z_s$.}
        \label{fig:spk_emb} 
        \vspace{2mm}
    \end{subfigure}
    ~
    \begin{subfigure}{\columnwidth}
        \includegraphics[width=\columnwidth,height=0.45\columnwidth]{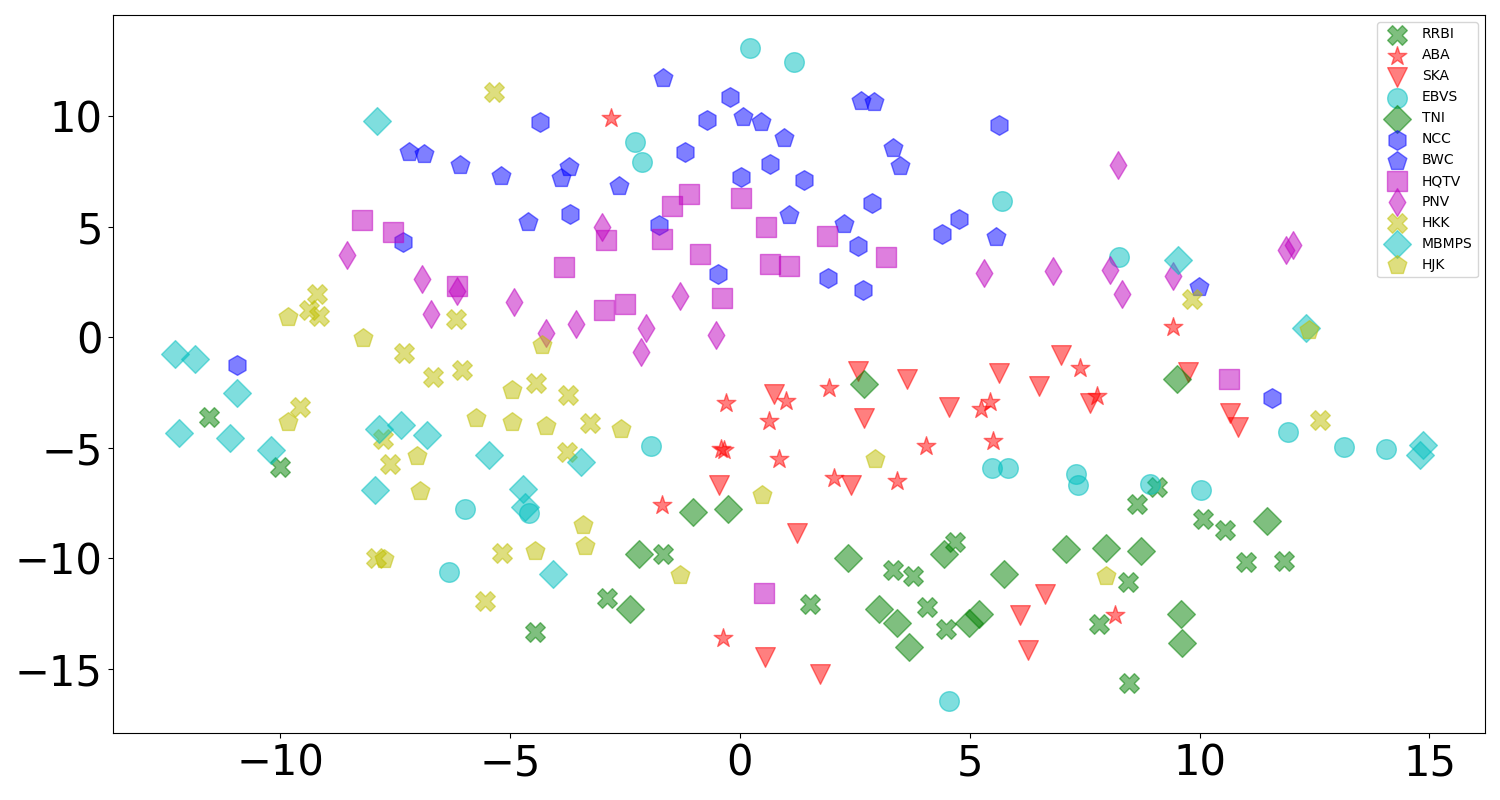}
        \vspace{-5mm}
        \caption{Accent embeddings $z_a$.}
        \label{fig:acc_emb}
        \vspace{2mm}
    \end{subfigure}
    ~
    \begin{subfigure}{\columnwidth}
        \includegraphics[width=\columnwidth,height=0.45\columnwidth]{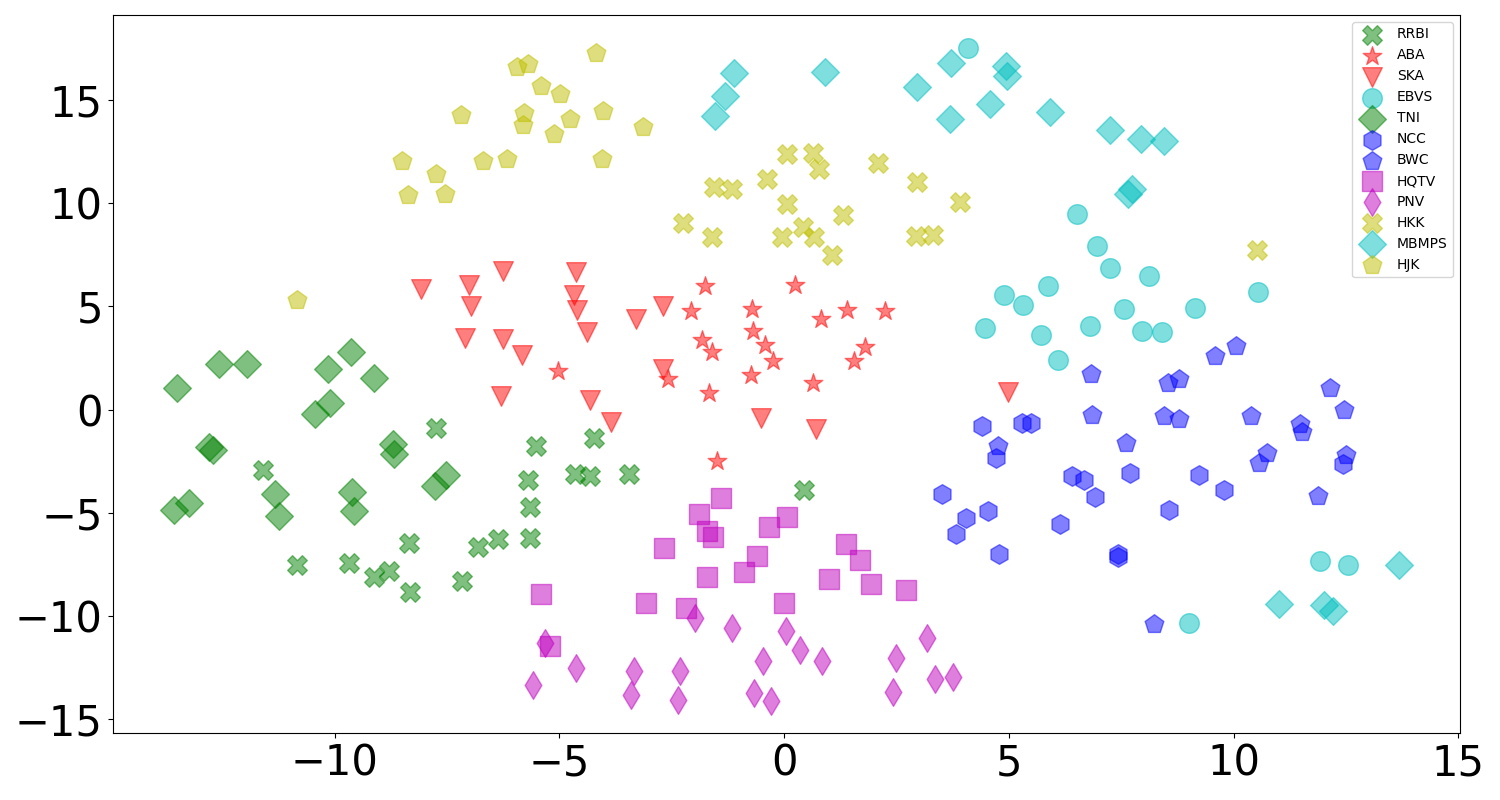}
        \vspace{-5mm}
        \caption{Combined embeddings of $z_s$ and $z_a$.}
        \label{fig:comb_emb}
    \end{subfigure}
    \vspace{-5mm}
    \caption{A t-SNE projection of the CVAE-NL embeddings. Each colour represents a different accent, whereas each shape represents a different speaker.}
    \vspace{-3mm}
\end{figure}

\subsection{Training and Inference}
We train all models with a batch size of 64 using the ADAM optimizer for 150k steps. The KL loss coefficient starts at $\num{1e-4}$ and increases linearly to $\num{5e-4}$ from 10k to 35k steps, then remains constant. Models are trained with target mel spectrograms as both reference input and target output, along with relevant text input.

During inference, the model can use reference audio for speaker and accent modeling. Using the VAE module's posterior encoder, we can sample from the two distributions to generate embeddings. For our experiments, we extract and store the $\mu_a$ and $\mu_s$ values from the validation set. We average these stored values and use them to represent specific speaker and accent, eliminating the need for further reference audio. In our baseline models, we average representations for each accent and speaker during inference. We also apply noise reduction to synthesized speech samples to improve sound quality and minimize signal distortion.

\begin{figure*}
    \centering
    \begin{subfigure}[t]{0.5\textwidth}
        \centering
        \includegraphics[height=1.5in]{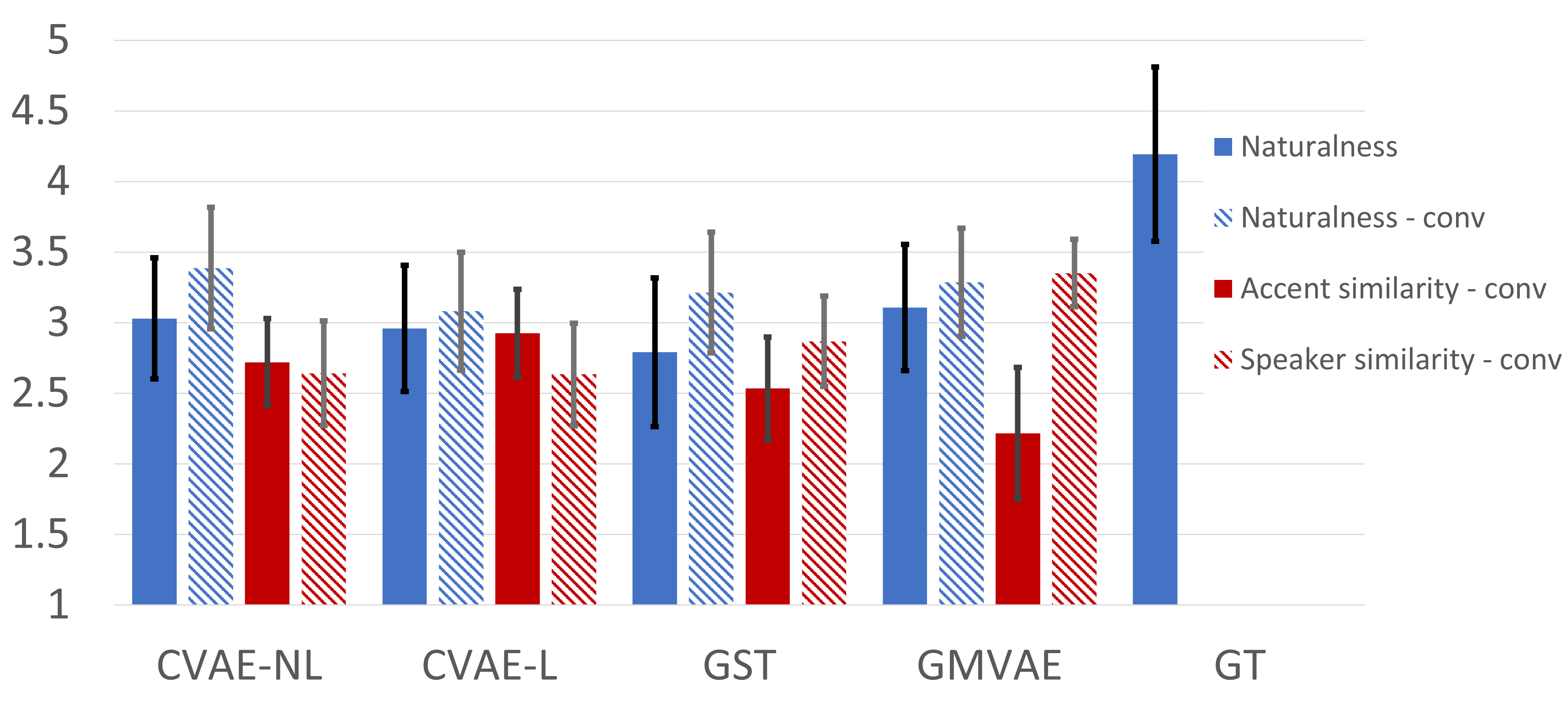} %
        \caption{MOS results with 95\% CI for naturalness without and with accent conversion; and accent and speaker similarity after accent conversion.}
        \label{fig:MOS}
    \end{subfigure}%
    ~ 
    \begin{subfigure}[t]{0.5\textwidth}
        \centering
        \includegraphics[height=1.2in]{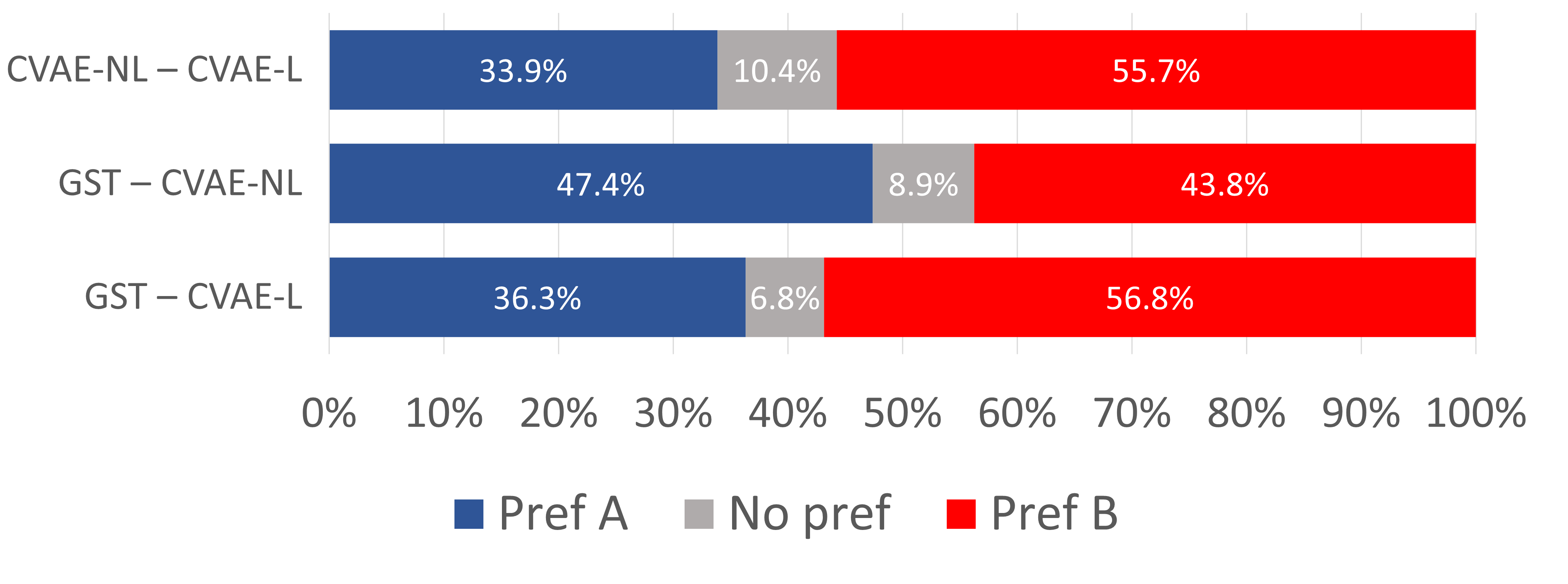}
        \caption{XAB general preference tests of accent-converted audio.}
        \label{fig:XAB}
    \end{subfigure}
    \vspace{-2mm}
    \caption{Subjective evaluation results.}
\vspace{-5mm}
\end{figure*}

\subsection{Accent and Speaker Modelling Analysis}
\label{subsec:analysis}

We visualized the embedding space of the CVAE-NL variant by encoding reference audio of the validation set and performing a t-distributed stochastic neighbor embedding analysis (t-SNE) for 12 of the 24 speakers. In Fig.~\ref{fig:spk_emb}, we can see that the speaker embeddings clearly form clusters per speaker. In Fig.~\ref{fig:acc_emb}, we observe overlap between speakers of the same specific accents. Interestingly, the combined (concatenated) embeddings in Fig.~\ref{fig:comb_emb} show even more compact clusters. This shows that the identity of each speaker is determined by both of the embeddings to a certain degree. One can imagine that if we move inside the combined embedding space by changing accent embeddings only, we might get a representation of a different speaker too. Naturally, we have observed this phenomenon in some of the synthesized audio samples with accent conversion, which influenced the design of the subjective evaluation tests, described in Section~\ref{subsec:sub_eval}.



\subsection{Objective Evaluation}
Mel Cepstral Distortion (MCD) \cite{mcd1} is used to evaluate the mel spectrogram reconstruction capabilities of each model, while Word Error Rate (WER) is used to evaluate the intelligibility of synthesized speech. The silero speech-to-text pre-trained models \cite{wer} are used for this purpose. The objective evaluation results presented in Table~\ref{tab:obj_eval} show that all models perform similarly. However, our proposed methods, CVAE-NL and CVAE-L, slightly outperform GST in terms of MCD but lag behind in terms of WER.
These results demonstrate that our proposed method achieves competitive speech quality.

\begin{table}[]
    \centering
    \caption{Objective Evaluation results (the lower, the better).}
    \resizebox{\columnwidth}{!}{%
    \begin{tabular}{@{}lccccc@{}}
    \toprule
    Metric & GT & CVAE-NL & CVAE-L & GMVAE & GST\\
    \midrule
    MCD $\downarrow$ &- &7.10 &7.176 & 7.375&7.502\\
    WER $\downarrow$ &0.1549&0.2311 &0.2008 &0.2228 &0.1959\\ \bottomrule
    \end{tabular}}
    \label{tab:obj_eval}
    \vspace{-5mm}
\end{table}
\vspace{-2mm}
\subsection{Subjective Evaluation}
\label{subsec:sub_eval}
We evaluate the proposed method and baselines through listening tests\footnote{Audio samples and code available via \rurl{dapwner.github.io/CVAE-Tacotron/}}. Sixteen participants listened to 192 samples each, assessing audio naturalness via Mean Opinion Score (MOS)~\cite{streijl2016mean}. Listeners rated samples from models in both non-conversion (original accent) and conversion (altered accent) settings. The naturalness results and other MOS tests are reported in Fig.~\ref{fig:MOS}. To assess statistical significance, we used a paired t-test to compare: 1) each non-conversion model, 2) each conversion model, and 3) each model's non-conversion result to its conversion counterpart.

In the non-conversion setting (1), GST's naturalness significantly differed ($p<0.05$) from GMVAE and CVAE-NL, with GST scoring the lowest. In the conversion setting (2), there was a significant difference between CVAE-NL and CVAE-L ($p<0.05$), suggesting CVAE-NL, which scored highest in naturalness after conversion, might be preferred. In the non-conversion to conversion setting (3), CVAE-NL and GST showed significant differences compared to their conversion counterparts ($p<0.001$), indicating that accent conversion increased perceived naturalness. For CVAE-L and GMVAE, the difference was not significant, indicating no loss in naturalness during accent conversion. All models' naturalness, with or without conversion, showed a highly significant difference compared to the ground truth ($p<0.001$), as expected. These results demonstrate that the proposed framework maintains naturalness during accent conversion.

We assessed the proposed method's performance in terms of accent and speaker similarity using MOS to quantify the perceived trade-off between accent and speaker identity after conversion. In the accent similarity test, listeners were given two reference samples: one of the source speaker (S1) to understand the original accent (A1) and one of the target accent (A2) represented by a different speaker (S2). They then rated the similarity of an audio sample of S1 in accent A2 to the reference accent A2. A paired t-test showed a statistically significant difference in accent similarity between CVAE-L and GST, and GMVAE (both with $p<0.001$). All models differed significantly in accent similarity (all $p<0.05$). The results indicate that the proposed methods outperform the baselines in accent conversion, with CVAE-L being the best performer, while the GMVAE baseline performed poorly.

In the speaker similarity, listeners rated how well the original speaker identity (S1 in A1) was retained after converting to the target accent (A2). The paired t-test showed a statistically significant difference between GMVAE and all other models ($p<0.001$), and between GST and all other models ($p<0.05$). While CVAE-L and CVAE-NL performed similarly to GST, they lagged behind GMVAE, reflecting the trade-off between retaining speaker identity and converting accent.
We conducted an XAB preference test to assess general preference for accent-converted audio without focusing on accent or speaker identity. Listeners were given a reference sample of the source speaker in their original accent and a reference sample of the target accent. They were asked: ``Imagine you are the original speaker J and you want your speech converted to a new accent K. Which sample, A or B, would you prefer for your new accent audio?'' This test aimed to evaluate performance and understand listener perception – whether they prefer more accented audio despite potential identity changes or prioritize maintaining identity. The GMVAE model was excluded due to its low conversion capability. Results (Fig.~\ref{fig:XAB}) show that GST and CVAE-NL models are equally preferred, while the CVAE-L model is favored over both, indicating it as the superior choice. This demonstrates that the proposed CVAE framework improves the state-of-the-art in accented TTS.

 


\subsection{Discussion on accent-identity balance}
CVAE shows promising results in both objective and subjective evaluations. Section \ref{subsec:analysis} highlights that while the embedding space supports strong accent conversion, it may disrupt speaker identity. Balancing accent and identity is challenging, as accent is a part of one's identity. The GMVAE captures speaker identity well but limits accent change. This issue is exacerbated by the dataset having only 4 speakers per accent as we can imagine a case of just 1 speaker per accent, where distinguishing between accent and speaker identity becomes impossible. In future work, we will focus on designing stronger mechanisms to further separate accent from speaker identity.

\section{Conclusion}
\label{sec:conc}
This paper introduces a novel framework for accented TTS, which fuses a Conditional VAE with the Tacotron2. The proposed framework allows for efficient synthesis of any chosen speaker's speech converted to any of the target accents. We conducted extensive objective and subjective tests to evaluate the efficacy of the proposed approach. The results show a strong performance in terms of the model's ability to synthesize natural-sounding speech in a converted accent, which pushes the state-of-the-art boundaries. We discuss the accent-identity balance and sketch out possible improvements in the development of accented TTS. Overall, the proposed framework has the potential to improve the quality and flexibility of TTS models and could play a significant role in the development of more advanced TTS systems.


\bibliographystyle{IEEEtran}
\bibliography{refs2,bibli}

\begin{thebibliography}{10}
\providecommand{\url}[1]{#1}
\csname url@samestyle\endcsname
\providecommand{\newblock}{\relax}
\providecommand{\bibinfo}[2]{#2}
\providecommand{\BIBentrySTDinterwordspacing}{\spaceskip=0pt\relax}
\providecommand{\BIBentryALTinterwordstretchfactor}{4}
\providecommand{\BIBentryALTinterwordspacing}{\spaceskip=\fontdimen2\font plus
\BIBentryALTinterwordstretchfactor\fontdimen3\font minus \fontdimen4\font\relax}
\providecommand{\BIBforeignlanguage}[2]{{%
\expandafter\ifx\csname l@#1\endcsname\relax
\typeout{** WARNING: IEEEtran.bst: No hyphenation pattern has been}%
\typeout{** loaded for the language `#1'. Using the pattern for}%
\typeout{** the default language instead.}%
\else
\language=\csname l@#1\endcsname
\fi
#2}}
\providecommand{\BIBdecl}{\relax}
\BIBdecl

\bibitem{wells1982accents}
J.~C. Wells and J.~C. Wells, \emph{Accents of English: Volume 1}.\hskip 1em plus 0.5em minus 0.4em\relax Cambridge University Press, 1982.

\bibitem{wang2017tacotron}
Y.~Wang, R.~Skerry-Ryan, D.~Stanton, Y.~Wu, R.~J. Weiss, N.~Jaitly, Z.~Yang, Y.~Xiao, Z.~Chen, S.~Bengio \emph{et~al.}, ``Tacotron: Towards end-to-end speech synthesis,'' \emph{arXiv preprint arXiv:1703.10135}, 2017.

\bibitem{shen2018natural}
J.~Shen, R.~Pang, R.~J. Weiss, M.~Schuster, N.~Jaitly, Z.~Yang, Z.~Chen, Y.~Zhang, Y.~Wang, R.~Skerrv-Ryan \emph{et~al.}, ``Natural tts synthesis by conditioning wavenet on mel spectrogram predictions,'' in \emph{IEEE ICASSP)}.\hskip 1em plus 0.5em minus 0.4em\relax IEEE, 2018, pp. 4779--4783.

\bibitem{ren2019fastspeech}
Y.~Ren, Y.~Ruan, X.~Tan, T.~Qin, S.~Zhao, Z.~Zhao, and T.-Y. Liu, ``Fastspeech: Fast, robust and controllable text to speech,'' \emph{Advances in Neural Information Processing Systems}, vol.~32, 2019.

\bibitem{ren2020fastspeech}
Y.~Ren, C.~Hu, X.~Tan, T.~Qin, S.~Zhao, Z.~Zhao, and T.-Y. Liu, ``Fastspeech 2: Fast and high-quality end-to-end text to speech,'' in \emph{International Conference on Learning Representations}, 2020.

\bibitem{wang2018style}
Y.~Wang, D.~Stanton, Y.~Zhang, R.-S. Ryan, E.~Battenberg, J.~Shor, Y.~Xiao, Y.~Jia, F.~Ren, and R.~A. Saurous, ``Style tokens: Unsupervised style modeling, control and transfer in end-to-end speech synthesis,'' in \emph{ICML}.\hskip 1em plus 0.5em minus 0.4em\relax PMLR, 2018, pp. 5180--5189.

\bibitem{hsu2018hierarchical}
W.-N. Hsu \emph{et~al.}, ``Hierarchical generative modeling for controllable speech synthesis,'' \emph{arXiv preprint arXiv:1810.07217}, 2018.

\bibitem{melechovsky2023learning}
J.~Melechovsky, A.~Mehrish, D.~Herremans, and B.~Sisman, ``Learning accent representation with multi-level vae towards controllable speech synthesis,'' in \emph{2022 IEEE Spoken Language Technology Workshop (SLT)}.\hskip 1em plus 0.5em minus 0.4em\relax IEEE, 2023, pp. 928--935.

\bibitem{oord2016wavenet}
A.~v.~d. Oord, S.~Dieleman, H.~Zen, K.~Simonyan, O.~Vinyals, A.~Graves, N.~Kalchbrenner, A.~Senior, and K.~Kavukcuoglu, ``Wavenet: A generative model for raw audio,'' \emph{arXiv preprint arXiv:1609.03499}, 2016.

\bibitem{sohn2015learning}
K.~Sohn, H.~Lee, and X.~Yan, ``Learning structured output representation using deep conditional generative models,'' \emph{NeurIPS}, vol.~28, 2015.

\bibitem{zhao2018l2}
G.~Zhao, S.~Sonsaat, A.~Silpachai, I.~Lucic, E.~Chukharev, J.~Levis, and R.~Gutierrez, ``L2-arctic: A non-native english speech corpus.'' in \emph{INTERSPEECH}, 2018, pp. 2783--2787.

\bibitem{mcd1}
R.~Kubichek, ``{Mel-cepstral distance measure for objective speech quality assessment},'' \emph{Communications, Computers and Signal Processing}, pp. 125--128, 1993.

\bibitem{wer}
``Silero models:pre-trained enterprise-grade stt/tts models and benchmarks,'' accessed: 2022-07-10.

\bibitem{streijl2016mean}
R.~C. Streijl, S.~Winkler, and D.~S. Hands, ``Mean opinion score (mos) revisited: methods and applications, limitations and alternatives,'' \emph{Multimedia Systems}, vol.~22, no.~2, pp. 213--227, 2016.

\end{thebibliography}

\end{document}